\documentclass[manuscript]{aastex}

\newcommand\simlt{\lower.5ex\hbox{$\; \buildrel < \over \sim \;$}}

\slugcomment{ApJ, to be submitted}

\shorttitle{"Leading blob" model for Mkn~501}
\shortauthors{E. Lefa et al.}

\begin{document}

\title{"Leading blob" model in a stochastic acceleration scenario: the case of the 2009 flare of Mkn~501}

\author{E. Lefa\altaffilmark{1,2}}
\affil{$^1$ Max-Planck-Institut f\"ur Kernphysik, P.O. Box 103980, 69029 Heidelberg, Germany}
\affil{$^2$ Landessternwarte, K\"onigstuhl 12, 69117 Heidelberg, Germany}
\email{eva.lefa@mpi-hd.mpg.de}
\author{F.A. Aharonian\altaffilmark{1,3}}
\affil{$^1$ Max-Planck-Institut f\"ur Kernphysik, P.O. Box 103980, 69029 Heidelberg, Germany}
\affil{$^3$ Dublin Institute for Advanced Studies, 31 Fitzwilliam Place, Dublin 2, Ireland}
\and
\author{F.M. Rieger\altaffilmark{1,4}}
\affil{$^1$ Max-Planck-Institut f\"ur Kernphysik, P.O. Box 103980, 69029 Heidelberg, Germany}
\affil{$^4$ European Associated Laboratory for Gamma-Ray Astronomy, jointly supported by CNRS and MPG}

\begin{abstract}
Evidence for very hard, intrinsic $\gamma$-ray source spectra, as inferred after correction for absorption
in the extragalactic background light (EBL), has interesting implications for the acceleration and radiation
mechanisms acting in blazars. A key issue so far has been the dependance of the hardness of the $\gamma$-ray
spectrum on different existing EBL models. The recent {\it Fermi} observations of Mkn~501 now provide additional
evidence for the presence of hard intrinsic $\gamma$-ray spectra independent of EBL uncertainties. Relativistic
Maxwellian-type electron energy distributions that are formed in stochastic acceleration scenarios offer a plausible
interpretation for such hard source spectra. Here we show that the combined emission from different components
with Maxwellian-type distributions could in principle also account for more softer and broader, power law-like
emission spectra. We introduce a "leading blob" scenario, applicable to active flaring episodes, when one (or few)
of these components become distinct over the "background" emission, producing hard spectral features and/or
hardening of the observed spectra. We show that this model can explain the peculiar high-energy characteristics
of Mkn~501 in 2009, with evidence for flaring activity and strong spectral hardening at the highest $\gamma$-ray
energies.
\end{abstract}


\section{Introduction}

Blazars are prominent $\gamma$-ray emitters with many peculiar features, the origin of which are currently
discussed intensively in the general context of their multi-wavelength properties. During the last years two
distinct features have attracted special attention: variability detected at very short timescales and the origin
of very hard, intrinsic $\gamma$-ray source spectra when absorption in the Extragalactic Background Light
(EBL) is taken into account. Although the observed TeV spectra of these sources are steep, their de-absorbed
(EBL-corrected) spectra appear intrinsically hard. Current uncertainties on the EBL flux level and spectrum
(cf. \citealt{primack11} for a recent review) introduce difficulties in defining how hard the absorption-corrected
source spectra are. However, in some characteristic cases, like for the distant blazars 1ES 1101-232 ($z=186$)
and 1ES 0229+200 ($z=0.139$), the emitted spectra still tends to be very hard, with an intrinsic photon index
$\Gamma\leq 1.5$, even when corrected for low EBL flux levels (\citealt{ahar06,ahar07}).

The recent {\it Fermi} detection of variable $\gamma$-ray emission from the nearby ($z=0.034$) TeV blazar
Mkn~501 in 2009 now removes this point of uncertainty, providing strong evidence for hard intrinsic $\gamma$-ray
source spectra independently of questions related to the EBL. As already indicated in the original {\it Fermi}
paper on Mkn~501 (\citealt{abdo2010a}), the spectrum above 10 GeV seems to become much harder during a
($\sim30$-day) flaring state. A recent, independent analysis of the same data by \cite{nero11} shows that the
(10 GeV-200 GeV) flare spectrum could be as hard as $\Gamma \simeq 1.1$. While the {\it Fermi} collaboration
did not comment much on the possible origin of the hard flare spectrum, \cite{nero11} put forward the hypothesis
that the hard spectrum flare could result from an electromagnetic cascade in the intergalactic medium, provided
that the strength of the intergalactic magnetic field is smaller than $10^{-16}$ G and that primary $\gamma$-rays
with 100 TeV can escape from the central compact region.

The noted small photon indices are generally not easy to achieve in standard leptonic scenarios, i.e.,
synchrotron self-Compton (SSC) or external Compton (EC) models, because radiative cooling tends to produce
particle energy distributions that are always steeper than $dN/dE\propto E^{-2}$, irrespective of the initially
injected particle spectrum. The corresponding TeV photon index would then be $\Gamma\geq 1.5$. Moreover,
as the emission from these objects peaks at very high energies where suppression of the cross-section due to
Klein-Nishina effects becomes important, even steeper intrinsic photon spectra are to be expected.

Hence, at first glance, hard $\gamma$-ray spectra might be somewhat easier to achieve in hadronic scenarios like the
proton synchrotron model. For a proton spectra of, e.g., $dN_{p}/dE=A E^{-s} \exp[-(E/E_0)^\beta]$ the $\gamma$-ray
spectrum would be $E_\gamma dN_{\gamma}/dE_\gamma \propto  \nu^{-(s-1)/2} \exp[-(\nu/\nu_0)^{\beta/\beta+2}]$. Thus at
energies below the synchrotron cutoff $h\nu_c \leq 300 \delta$ GeV (\citealt{aharonian00}) very hard $\gamma$-ray
spectra may occur. For the magnetic field $B \sim 100$~G required in such models, the protons responsible for
$\gamma$-rays below $h\nu_c$ are not effectively cooled. Hence, for $s \leq 2$, the $\gamma$-ray spectrum
could be harder than $\nu F_\nu \propto \nu^{1/2}$, with the hardest possible $\nu F_\nu \propto \nu^{4/3}$ in
the case of pile-up type proton distributions that might be realized in, e.g., converter-type particle acceleration
mechanisms (\citealt{derishev03}).

On the other hand, in the context of leptonic models we know that very narrow electron distributions are able to produce hard
$\gamma$-ray source spectra. These can be either power-law distributions with large value of the minimum
cut-off, provided the magnetic field is sufficiently small to avoid radiative losses (\citealt{katar07,tavecchio09}), or
provided adiabatic losses dominate (\citealt{lefa11}). Alternatively, relativistic Maxwell-like distributions formed
by a stochastic acceleration process that is balanced by radiative losses can be a viable option (\citealt{lefa11}).
In both cases VHE spectra as hard as $E_{\gamma}dN/dE_{\gamma} \propto E^{1/3}_{\gamma}$ for the SSC
case, and $E_{\gamma}dN/dE_{\gamma} \propto E_{\gamma}$ for EC models can be generated. Maxwellian-type
particle distributions are especially attractive for the interpretation of the inferred hard $\gamma$-ray source
spectra because they resemble to some extent mono-energetic distributions (the hardest possible injection
spectra). An important question that arises however then is, whether such distributions can also account for more
softer and broader photon spectra.

Interestingly, the hard high-energy component of Mkn~501 emerged in a flaring state in May 2009, during which
little flaring activity was detected at energies below 10 GeV. This suggests that the flare is related to an emission
zone that is confined both in space (compact) and time (short, on a timescale of $\sim 30$ days). Additionally,
there is no evidence for a similar, simultaneous increase at X-ray energies (\citealt{abdo2010a}).

Here we propose a leptonic multi-zone scenario that can accommodate softer emission spectra as well as flaring
episodes with hard spectral features, like the one observed in Mkn~501. In this scenario, the observed radiation
comes from several emitting regions ("blobs"), in which electrons are accelerated to relativistic energies through
stochastic acceleration, forming pile-up distributions. All blobs are considered to have similar parameters apart
from the characteristic energies ("temperatures") of their Maxwell-like distributions. We note that a "multi-blob"
scenario, with power-law components characterized by different trajectories (viewing angles ) has also been
proposed in the past to account for the TeV emission from non-aligned AGNs \citep{lenain08}. What distinguishes
the model introduced here is that, due to acceleration and losses, each component is considered to only carry
a narrow (pile-up) particle distribution, with broad spectra being formed by an ensemble of components. We
show, for example, that in the case where the total energy of the particles is the same for all blobs, the combined
emission leads to a spectrum very similar to the one arising from a power-law distribution $dN/dE\propto E^{-p}$
with index $p=2$. On the other hand, distinct spectral feature can appear once a leading zone dominates. This
could happen, for example, if (a) the value of the temperatures changes from blob to blob, and/or if (b) the
energetics of a single blob changes.

\section{A multi-zone scenario}
Let us consider $N$ regions in which electrons are stochastically accelerated (e.g., by scattering off randomly moving
Alfv\'en waves) up to energies where acceleration is balanced by synchrotron or inverse Compton (Thomson) losses.
Their steady-state energy distributions $n(\gamma) \propto \gamma^2 f(\gamma)$ then take on a relativistic
Maxwellian-type form (\citealt{Schlickeiser85,aharonian86})
\begin{equation}\label{max}
n_{i}(\gamma)=A_{i}\gamma^{2}e^{-\left(\frac{\gamma}{\gamma_{c_{i}}}\right)^{b}}\,,
\end{equation}
($b \neq 0$; $i=1,..,N$) with exponential cut-off Lorentz factor
\begin{equation}\label{gcut}
\gamma_{c_i}=\left(\frac{b D_0}{\beta_s}\right)^{1/b} (m_e c^2)^{-1}\,,
\end{equation} and normalization factor $A_i$. Here, $\beta_s$ refers to energy losses due to synchrotron radiation
given by
\begin{equation}
\frac{dp}{dt}=-\beta_{s}
p^{2}=-\frac{4}{3}(\sigma_{T}/m_e^2c^2) U_{B}p^{2}\,,
\end{equation}
and the constant $D_0$ is given by the diffusion coefficient $D_p=\frac{p^{2}}{3\tau}(\frac{V_{A}}{c})^{2}\equiv D_0
p^{1-b}$, with $V_{A}=\frac{B}{\sqrt{4\pi \rho}}$ the Alfv\'{e}n speed and $\tau =\lambda/c\propto p^{b-1}$, $b \geq 1$,
the mean scattering time (\citealt{lefa11}). The shape of the exponential cut-off (characterized by the parameter $b$)
is related to the turbulence wave spectrum $W(k) \propto k^{-q}$ as $b=q$. Note that if the particle distributions would
be shaped by inverse Compton cooling in the Klein-Nishina regime, a smoother exponential cut-off is expected
(e.g., \citealt{stawarz08}).

For simplicity, we consider below the situation where all blobs have similar properties (e.g., magnetic field strength,
linear size, Doppler factor) but different values for the characteristic energy $\gamma_{c_i}$. In the case of synchrotron
losses and scattering off Alfv\'en waves, $\gamma_{c_i}$ depends on the magnetic field and the bulk density of the flow.
Thus, the modification in $\gamma_{c_i}$ that we assume for each blob might be related to a non-homogeneous bulk
flow.\\
The total energy density $E_i$ that the relativistic particles gain through scattering off Alfv\'en waves is calculated to be
\begin{equation}\label{energy}
E_i=\int^{\infty}_{1} \gamma n_{i}(\gamma)d\gamma m_e c^2
      =A_{i}m_e c^2 \frac{\gamma^{4}_{c_i}}{b}\Gamma[\frac{4}{b},\frac{1}{\gamma^{b}_{c_i}}]\,,
\end{equation}
where $\Gamma[a,z]$ is the incomplete $\Gamma$ function. We can now express the normalization factor as a function
of the "temperature" ($\gamma_{c_i}$) and the energy $E_i$,
\begin{equation}
A_{i}=\frac{E_i b}{m_e c^2\gamma^{4}_{c_i}\Gamma[\frac{4}{b},\frac{1}{\gamma^{b}_{c_i}}]}\,,
\end{equation}
which decreases as the cut-off energy increases, with a dependency well approximated by $A_{i}\propto\gamma^{-4}_{c}$
for all values of the coefficient $b$ of interest. This is easily seen if we change the lower limit of the previous integration to
$0$, in which case the result becomes
\begin{equation}\label{norm}
A_{i} = \frac{4 E_i}{m_e c^2\gamma^{4}_{c_{i}}\Gamma[(4+b)/b]}\,,
\end{equation} with $\Gamma[z]$ denoting the $\Gamma$ function.

The combination of the above electron distributions can lead to power-law-like particle distributions if the temperatures of
the different components do not differ significantly and if the components contribute equally to the overall spectra. The
power-law index then essentially depends on the amount of energy given to the non-thermal particles in each blob, i.e.,
on how the total energy $E_i$ of each component scales with temperature ($\gamma_{c_i}$). Steep spectra may arise
if, e.g., the low temperature components dominate whereas harder spectra may occur if more energy is contained in the
high-temperature blobs.

\begin{figure}[!h]
\epsscale{0.1pt}
\begin{center}
\includegraphics[width=300pt,angle=-90]{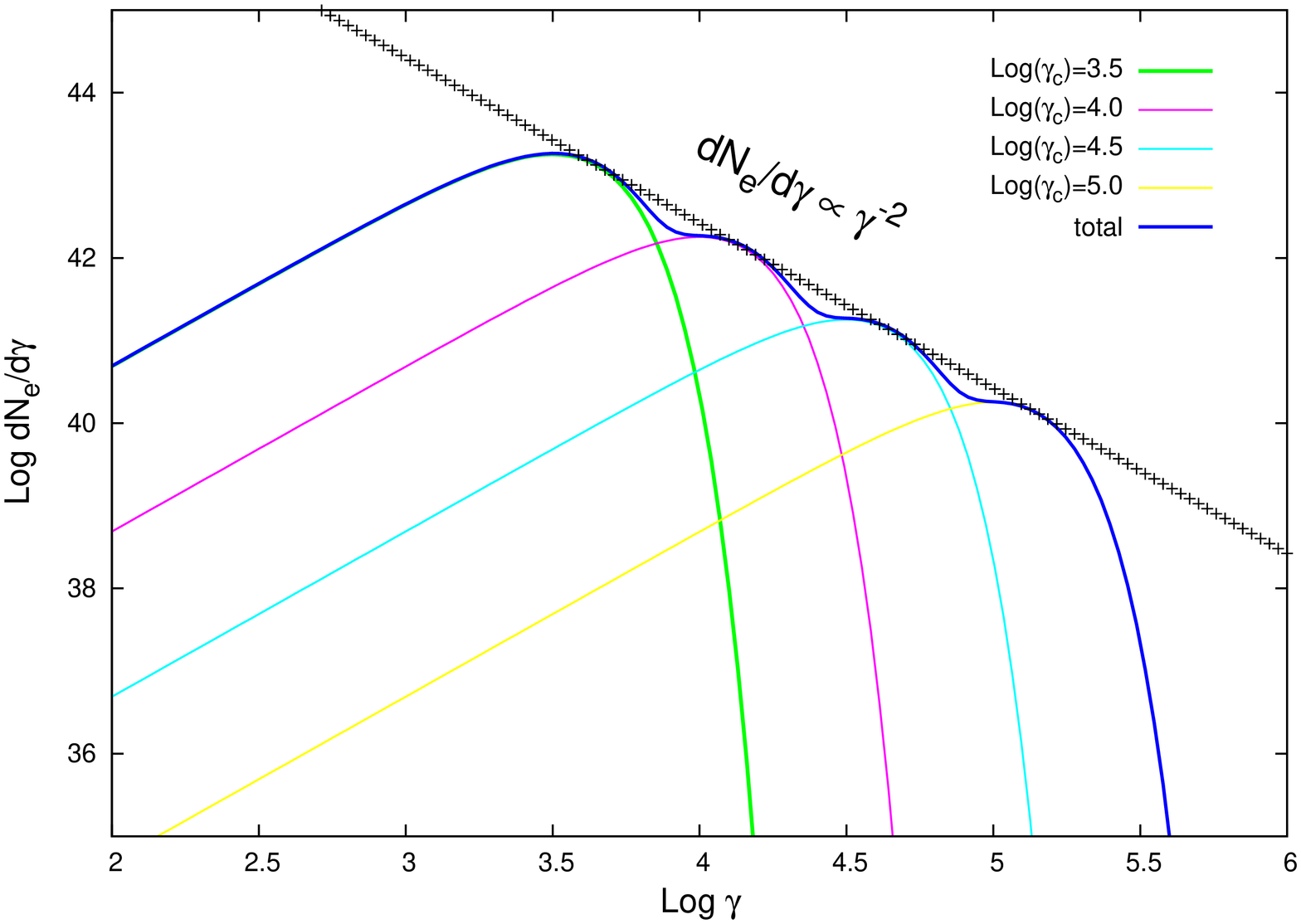}
\caption{The total electron energy distribution from $N=4$ blobs with different temperatures ($\gamma_c$) but the same
total energy $E_i$ tends to establish a power-law $dN_{e}/d\gamma\propto\gamma^{-2}$ between the minimum and
maximum characteristic energies $\gamma_{c}$. Below $\gamma_{\rm c,min}$, $dN_{e}/d\gamma\propto\gamma^{2}$,
whereas for $\gamma>\gamma_{\rm c,max}$ one has $dN_{e}/d\gamma\propto e^{-(\gamma/\gamma_{\rm c,max})^2}$.
Here, the exponential cut-off index is $b=2$ and other parameters used are $B=0.1$ G and $E_i=2 \times 10^{44}$ erg.}\label{MKNtheoele}
\end{center}
\end{figure}

In Fig.~\ref{MKNtheoele} an example for the total differential electron number is shown assuming $E_i=$ constant.
For the plot, an exponential cut-off index $b=2$ and $N=4$ have been chosen. The temperatures are equally spaced
on logarithmic scale. Then, the total energy distribution approximately forms a power-law $dN_{e}/d\gamma \propto
\gamma^{-s}$ with index $s \simeq 2$ between the minimum and maximum temperatures. As discussed above, this
"special" value of the power-law index results from the assumption that all blobs have the same total energy, so that
 $\gamma^{2} dN_{e}/d\gamma$ is the same for each zone.
This can be demonstrated more formally by looking for the "envelope", i.e., the mathematical function that describes
the curve which is tangent to each of the curves $n_i$ in the $(dN_e/d\gamma,\gamma_c)$ plane at some point. This
function approximates the sum of the energy distribution of the different components as $N>>1$ and gives the
characteristic behavior of the total distribution above the minimum and below the maximum temperature. It can be
found by solving the set of equations
\begin{equation}
F(\gamma,\gamma_c)=0,\;\;\;\;\vartheta_{\gamma_c}F(\gamma,\gamma_c)=0\,,
\end{equation}
where $F(\gamma,\gamma_c)\equiv dN_e/d\gamma - A(\gamma_c)~\gamma^2 \exp[-(\gamma/\gamma_c)^b]$.
If we assume that the energy in non-thermal particles is the same for each component, then we find
\begin{equation}
dN_e/d\gamma=c'(b)~\gamma^{-2}, \;\;\;\;\gamma_{\rm c,min}<\gamma<\gamma_{\rm c,max}\,,
\end{equation}
where $c'(b)=4(4/b)^{4/b}e^{-b/4}/m_ec^2\Gamma[\frac{4+b}{b}]$ (cf. Fig.~\ref{MKNtheoele}).
On the other hand, if particle acceleration to higher energies goes along with a decrease in total energy, e.g.,
$E\propto 1/\gamma_{c}$, then steeper power law spectra can appear, i.e., $\frac{dN_{e}}{d\gamma}\propto
\gamma^{-3}$. Note that in all these cases, radiative cooling is already taken into account.

Note that changing the total energy with $\gamma_{c_i}$ in each component could also be interpreted as
changing the number of contributing blobs as a function of $\gamma_{c_i}$, assuming that each component
has the same energy. This could be formally accommodated by introducing a statistical weight $w_i$, so that
the overall spectrum is expressed as
\begin{equation}
dN_{e}/d\gamma=\sum_{i=1}^{i=N} w_i n_i(\gamma)\,.
\end{equation}
Harder spectra may then occur, for example, if more blobs with higher temperatures exist and vice versa.
Hence, a conclusion similar to the above can be drawn, once the statistical weights vary correspondingly
with temperature ($\gamma_{c_i}$). In a continuous analogue, we may write
\begin{equation}
dN_{e}/d\gamma=\sum_{i=1}^{i=N} w_i n_i(\gamma)\rightarrow\int^{T_N}_{T_1}W(T)n(\gamma,T)dT
\end{equation}
where $W(T)$ is the spectrum of the number of components per temperature. Since Maxwellian-type electron
distributions behave, to some extent, like mono-energetic ones, the total energy distribution between the minimum
and the maximum temperature mimics the spectrum of the number of the blobs $W(T)$ (provided $W(T)$ does
not rise quicker than $T^2$). Thus, in principle a variety of spectra may arise, depending on the choice of $W(T)$.
Conversely, observations of extended power law-like energy distributions then impose constraints on how $W(T)$
of a source can vary with temperature.

The SSC spectrum, arising as the sum of the different Maxwell-like distributions of Fig.~\ref{MKNtheoele}, is shown
in Fig.~\ref{MKNtheo}. The synchrotron flux resembles the flux that would be emitted by a power-law particle
distribution of index $2$. Between the frequencies related to the minimum and maximum temperatures, $\nu_{\rm min}
\propto B\gamma^{2}_{\rm c,min}$ and $\nu_{\rm max}\propto B\gamma^{2}_{\rm c,max}$ it exhibits a $F_{\nu}\propto
\nu^{-1/2}$ behavior. For $\nu<\nu_{\rm min}$ it follows the characteristic $F_{\nu}\propto\nu^{1/3}$ synchrotron
emissivity function, while for $\nu \geq \nu_{\rm max}$ the exponential cut-off becomes smoother $F_{\nu}\propto
\exp{[-(\nu/\nu_{\rm max})^\frac{b}{2+b}}]$ (\citealt{frit89,zirak07}).

In the present model, the electrons in each blob are considered to only up-scatter their own synchrotron photons
and not the ones emitted from the other blobs. The photon fields produced by the other components therefore do
not contribute to the emitted Compton spectrum of each blob. In the Thomson regime then the up-scattered photon
spectrum again approaches a power-law behavior similar to the synchrotron one, i.e. $F_{\nu} \propto \nu^{-1/2}$.
Once Klein-Nishina effects become important, the suppression of the cross-section makes the high-energy spectrum
steeper, as expected. In the case of discrete zones, where the particle distributions and synchrotron photons are
almost mono-energetic, this happens at (intrinsic) energies greater than $\gamma_{c_i}(B/B_{cr})\gamma^{2}_{c_i}
>1$, where $B_{cr}=m_e^2c^3/ (e\hbar)$. Here, $(B/B_{cr})\gamma^{2}_{c_i}$ is the peak energy of the synchrotron
photons emitted by the i-th blob with temperature $\gamma_{c_i}$. Below $\nu\propto \gamma^{2}_{c,min}B
\gamma^{2}_{c,min}$ the Compton flux reveals the characteristic $1/3$ slope, reflecting the low-energy synchrotron
spectrum. In the Klein-Nishina regime, the exponential VHE cut-off mimics the shape of the electron cut-off,
$F_{\nu}\propto\exp{[-(\nu/\nu_{\rm max})^b}]$, and is steeper compared to the synchrotron spectrum.

\section{The origin of hard $\gamma$-ray flare spectra}
Once a single component becomes dominant in the overall emission, as naturally expected for a flaring stage,
hard spectral features can arise. This is more evident in the Compton part of the spectrum as (in the Thomson
regime) the separation of the VHE peaks scales as $\sim\gamma^{4}_{c}$ and is greater than in the synchrotron
case ($\sim\gamma^{2}_{c}$). The energetics of such a leading component, which is responsible for an observed
flare, could change for different reasons: The total (intrinsic) energy offered to the accelerated particles or/and
the temperature of the distribution could increase, for example, due to changes in the bulk flow properties or
due to an increased injection of seed particles that undergo stochastic acceleration. Another possibility concerns
an increase of the Doppler factor. Already a slight change of the line-of-sight angle to the observer during the
propagation of a blob, could lead to the observation of a hard flare without an accompanying change of the
intrinsic energetics of the components.

The aforementioned possibilities can be applied to explain the high-energy flare of Mkn~501 observed in 2009.
To illustrate this, we assume for the "low state" that the total energy of each component drops as the temperature
increases in order to account for a steeper than $1.5$ spectrum. The scale assumed is $E\propto\gamma_c^{-1/4}$.
For the flaring state the normalization of the two components with the highest temperatures is increased by a factor
of $\sim 2$. Their temperature is also slightly increased. This leads to a hard high-energy flare above 10 GeV. Below
10 GeV the emission stays almost constant. Other parameters are kept the same for all blobs (i.e., magnetic field
$B=0.1$ G, blob radius $R=10^{14}$ cm and Doppler factor $\delta=30$). For this set of parameters, the synchrotron
flux is lower than the one observed, implying that X-ray emission would have to come from a different part of the jet.
This is consistent with the fact that little flux variation has been observed in X-rays.

\section{Conclusion}
Narrow energetic electron distributions, like relativistic Maxwellian-type ones, can successfully explain the very hard intrinsic
$\gamma$-ray spectra that arise in some sources once EBL absorption is taken into account \citep{lefa11}.

Here we have demonstrated that the superposition of emission from such distributions could also accommodate more softer
and broader $\gamma$-ray spectra. To show this, a multi-zone scenario was considered in which particles are accelerated through a stochastic acceleration process balanced by radiative (synchrotron/Thomson) losses in multiple zones characterized
by different temperatures (i.e., achievable maximum electron energies). For the parameters examined here, particle escape
can be neglected, and the particle distribution in each zone essentially takes a Maxwellian-type shape. Under reasonable conditions, the resultant overall (combined) particle energy spectra then approaches a power-law particle distribution
$dN_e/d\gamma \propto \gamma^{-s}$ over the energy range corresponding to the lowest and the highest temperature, with power index $s$ only depending on how the total energy (in non-thermal particles) in each component scales with temperature
(cut-off Lorentz factor $\gamma_c$). In the case where all parameters, apart from the temperature, are kept constant (in
particular the total energy in each component), the resultant power index approximates $s \rightarrow 2$. For similar
magnetic fields and Doppler factors, softer/harder $\gamma$-ray spectra could arise when the lower/higher temperature components dominate.

In this scheme, the dominance of one (or a few) of the radiating components could lead to a flaring state during which
hard spectral feature become apparent. This leading component might increase its luminosity for different reasons, e.g.,
due to a change of the Doppler factor or the injected energy. As shown above, such a scenario can account for the 2009
flare in Mkn~501 during which a strong hardening of the emission spectra above 100 GeV has been observed (\citealt{abdo2010a,nero11}. Mkn~501 is indeed known to be a source where detailed temporal and spectral modeling has
provided evidence for the contributions of different components (such as a steady X-ray component plus a variable SSC component, see \citealt{kraw02}).

While in the present work an SSC approach has been employed, similar features are to be expected in external Compton scenarios. In the latter case, an even stronger spectral hardening up to $F_{\nu}\propto\nu$ may occur, while in the SSC
case this is limited to $F_{\nu}\propto\nu^{1/3}$ \citep{lefa11}.

In summary, the "leading blob" model introduced here can explain in a natural way both the quiescent and flaring state in
Mkn~501. It will be interesting to check to what extent this also applies to other sources.

{}
\newpage

\begin{figure}[!h]
\epsscale{0.1pt}
\begin{center}
\includegraphics[width=300pt,angle=-90]{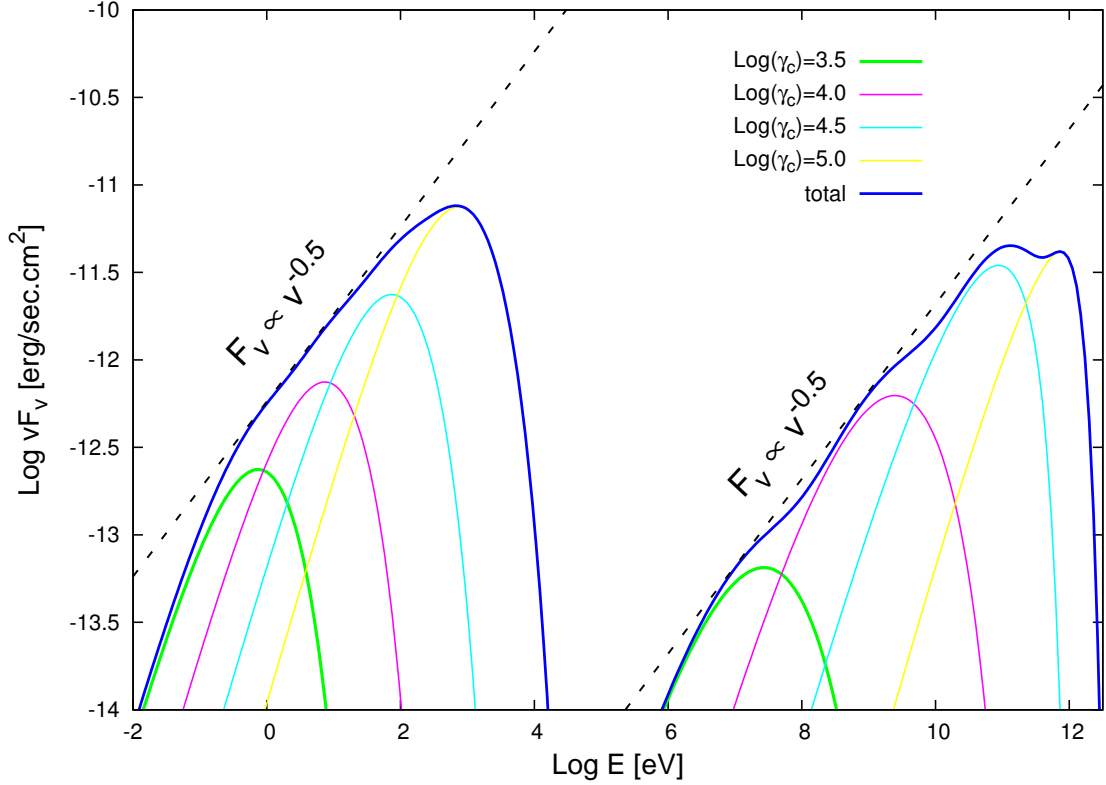}
\caption{Resultant SSC emission from the combination of different components with Maxwellian-like electron distributions. The
synchrotron flux exhibits a power-law behavior, $F_{\nu}\propto\nu^{-1/2}$, approximately between the energies related to the
minimum and maximum temperature. The same holds for the Compton flux in the Thomson regime, while in the Klein-Nishina
regime the spectrum becomes steeper. Doppler factor $\delta=30$ and cut-off index $b=2$ have been used. Other parameters
are $R=3\times10^{14}$ cm, $B=0.1$ G, $E_i=2\times10^{44}$ erg.}\label{MKNtheo}
\end{center}
\end{figure}\label{MKNtheo}

\begin{figure}[!h]
\epsscale{0.1pt}
\begin{center}
\includegraphics[width=300pt,angle=-90]{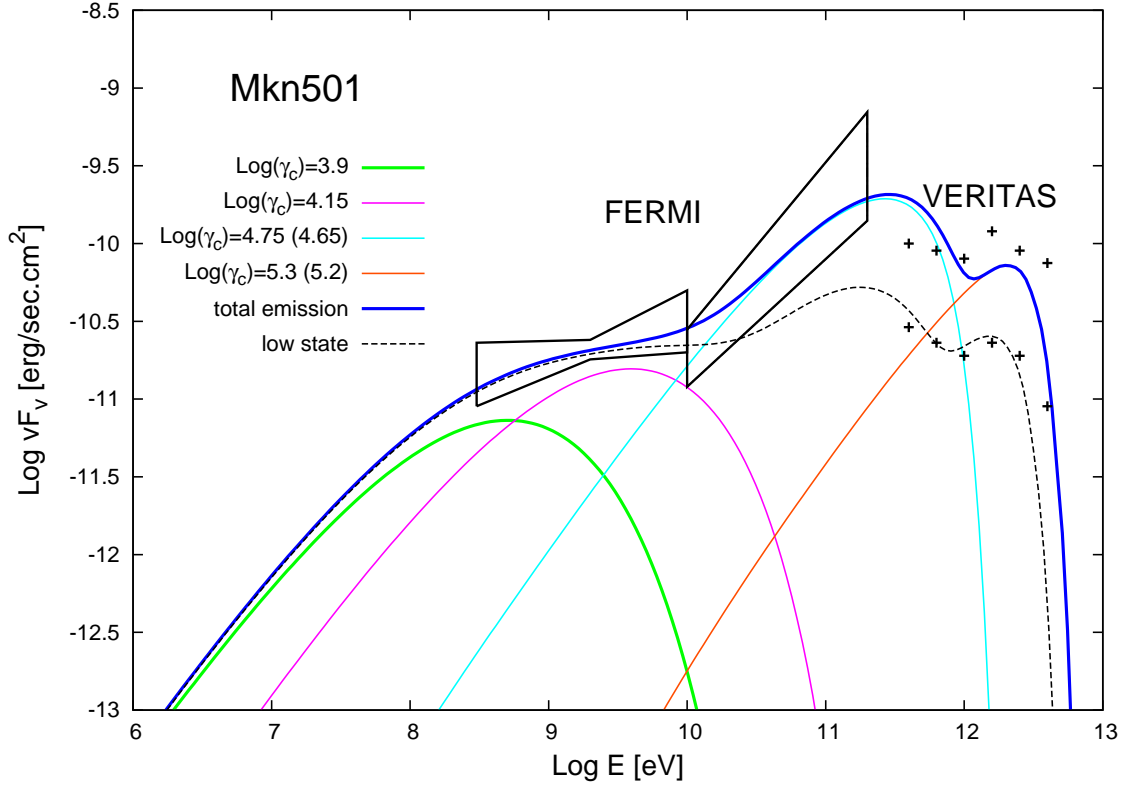}
\caption{Resultant SSC emission from the sum of different ($N=4$) blobs for the "low" (dashed-line) and "flaring state"
(blue line) of Mkn~501. The total energy given to the particles scales as $E_i\propto \gamma^{-1/4}_{c_i}$ and for the
low state the temperatures are $\log(\gamma_{c})=3.9,4.14,4.6$ and $5.2$, respectively. Parameters kept constant are
magnetic field $B=0.1$ G, blob radius $R=10^{14}$ cm, cut-off index $b=3$ and Doppler factor $\delta=30$. For the
flaring state, the two blobs with highest temperatures are assumed to be enhanced by a factor of $\sim2$ with their
temperatures slightly increased (to 4.74 and 5.3, respectively). Below 10 GeV, the flux is almost constant with respect
to the low state. For data points, see \citealt{abdo2010a} and \citealt{nero11}.}\label{MKN1}
\end{center}
\end{figure}

\begin{figure}[!h]
\epsscale{0.1pt}
\begin{center}
\includegraphics[width=300pt,angle=-90]{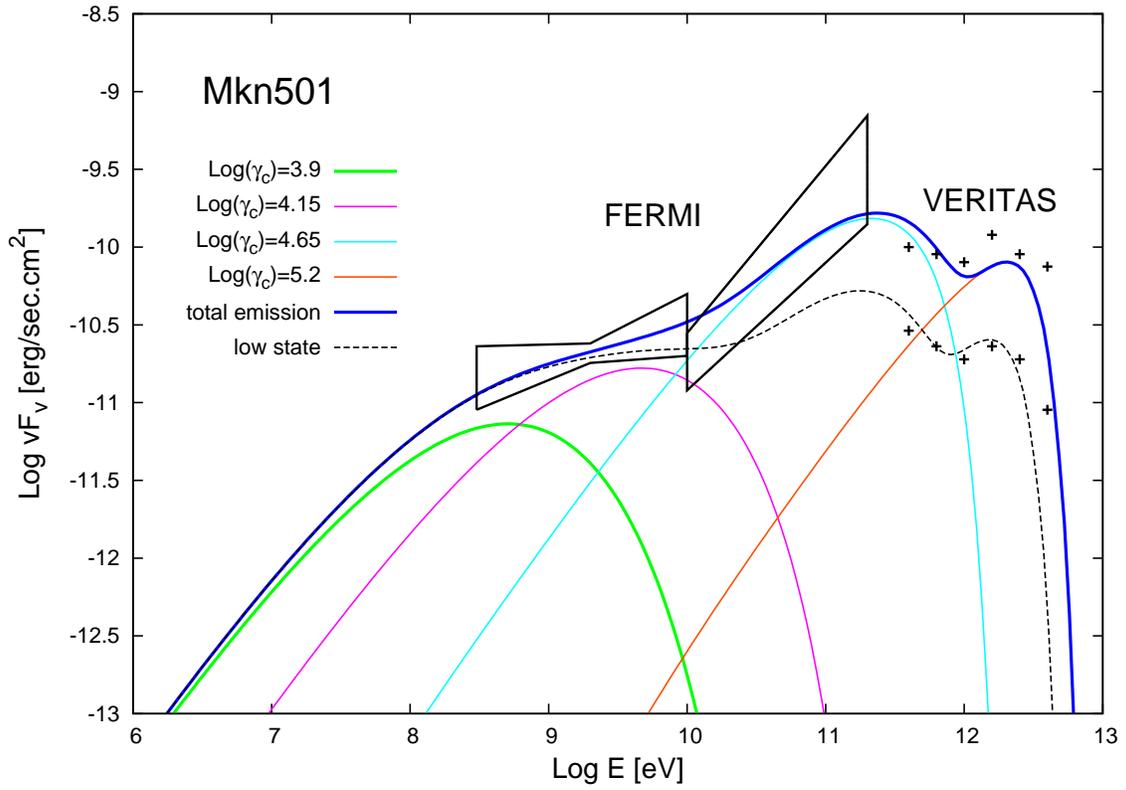}
\caption{Same as figure \ref{MKN1} but assuming the flaring state to occur due to a change of the Doppler factor of
the two components with the highest temperatures from $\delta=30$ to $\delta=40$.}\label{MKNdoppler}
\end{center}
\end{figure}

\begin{figure}[!h]
\epsscale{0.1pt}
\begin{center}
\includegraphics[width=300pt,angle=-90]{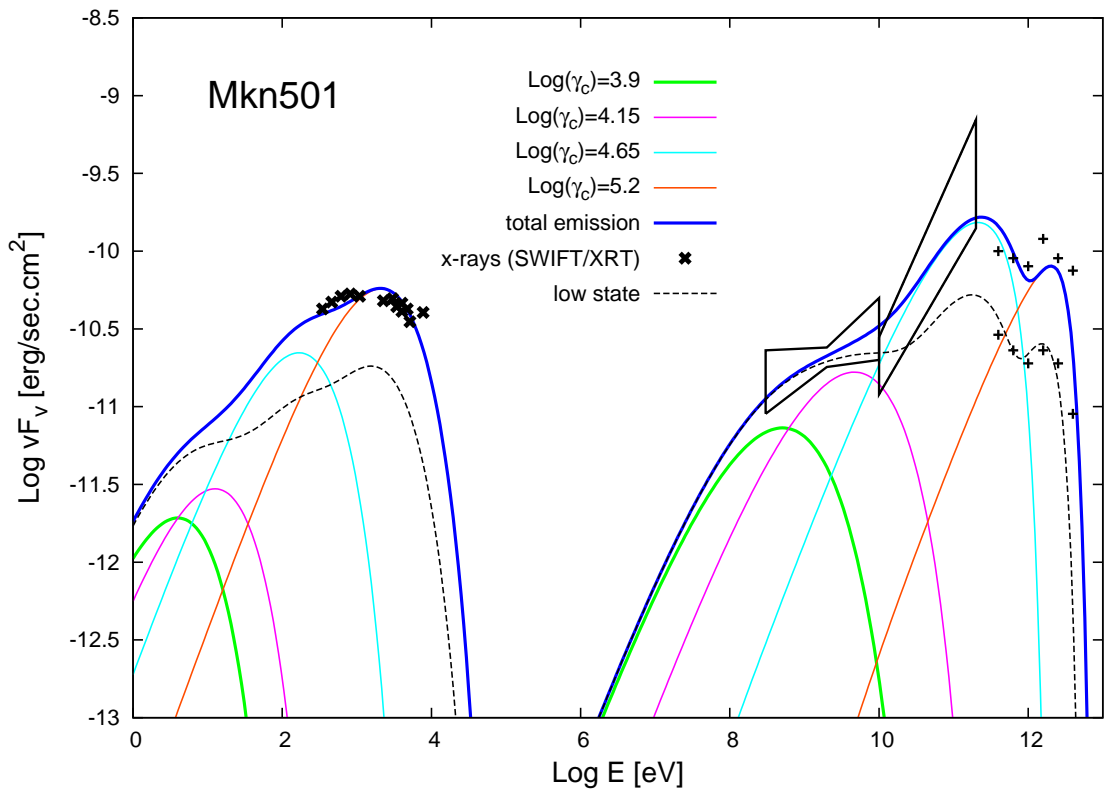}
\caption{Same as figure \ref{MKNdoppler} with the synchrotron part of the spectrum included. The X-ray regime is
considered to be dominated by emission from a different part of the jet. Thus, during the flare no/little variability at
X-ray energies would be observed as long as the synchrotron contribution from the "flaring" components does
not exceed the measured X-ray data.}\label{xrays}
\end{center}
\end{figure}


\begin{thebibliography}{}

\bibitem[\protect\citeauthoryear{Abdo et al.}{2011}]{abdo2010a}
Abdo, A.A. et al. 2011, ApJ, 727, 129

\bibitem[\protect\citeauthoryear{Aharonian et al.}{1986}]{aharonian86}
Aharonian, F.A., Atoyan, A. M., $\&$ Nahapetian, A. 1986, A$\&$A, 162, L1

\bibitem[\protect\citeauthoryear{Aharonian}{2000}]{aharonian00}
Aharonian, F.A. 2000, NewA, 5, 377

\bibitem[\protect\citeauthoryear{Aharonian et al.}{2006}]{ahar06}
Aharonian, F., et al. 2006, Nature, 440, 1081

\bibitem[\protect\citeauthoryear{Aharonian et al.}{2007}]{ahar07}
Aharonian, F., et al.  2007, A\&A, 475, L9

\bibitem[\protect\citeauthoryear{Derishev et al.}{2003}]{derishev03}
Derishev, E.V., Aharonian, F.A., Kocharovsky, V.V., Kocharovsky, Vl. V. 2003, Phys.
Rev. D, 68, 043003

\bibitem[\protect\citeauthoryear{Fritz}{1989}]{frit89}
Fritz, K.D. 1989, A\&A, 214, 14

\bibitem[\protect\citeauthoryear{Katarzy{\'n}ski et al.}{2007}]{katar07}
Katarzy{\'n}ski, K., Ghisellini, G., Tavecchio, F., Gracia, J., $\&$ Maraschi, L. 2006, MNRAS, 368, L52

\bibitem[\protect\citeauthoryear{Krawczynski et al.}{2002}]{kraw02}
Krawczynski, H., Coppi, C.S., \& Aharonian, F. 2002, MNRAS, 336, 721

\bibitem[\protect\citeauthoryear{Lefa et al.}{2011}]{lefa11}
Lefa, E., Rieger, F.M., \& Aharonian, F.A. 2011, ApJ, in press (doi:10.1088/0004-637X/737/1/1)

\bibitem[\protect\citeauthoryear{Lenain et al.}{2008}]{lenain08}
Lenain, J.-P., Boisson, C., Sol, H. \& Katarzy{\'n}ski, K., 2008, A\&A 478, 111

\bibitem[\protect\citeauthoryear{Neronov et al.}{2011}]{nero11}
Neronov, A., Semikoz, D. \& A.M. Taylor 2011, A\&A submitted (arXiv:1104.2801)

\bibitem[\protect\citeauthoryear{Primack et al.}{2011}]{primack11}
Primack, J. R. et al. 2011, Proc. of the 25th Texas Symposium (eds. F.A. Aharonian, W. Hofmann, F.M. Rieger),
AIP Conf. Proc. 1381, in press (arXiv:1107.2566)

\bibitem[\protect\citeauthoryear{Schlickeiser}{1985}]{Schlickeiser85}
Schlickeiser, R. 1985, A$\&$A, 143, 431

\bibitem[\protect\citeauthoryear{Stawarz \& Petrosian}{2008}]{stawarz08}
Stawarz, L., \& Petrosian, V. 2008, ApJ 681, 1725

\bibitem[\protect\citeauthoryear{Tavecchio et al.}{2009}]{tavecchio09}
Tavecchio, F., Ghisellini, G., Ghirlanda, G., Costamante, L., $\&$ Franceschini, A. 2009, MNRAS, 399, L59

\bibitem[\protect\citeauthoryear{Zirakashvili \& Aharonian}{2007}]{zirak07}
Zirakashvili, V.N., \& Aharonian, F. 2007, A\&A, 465, 695
\end{thebibliography}
\end{document}